# Causal Wave Mechanics and the Advent of Complexity.
# IV. Dynamical origin of quantum indeterminacy and wave reduction

A.P. Kirilyuk*

Institute of Metal Physics, Kiev, Ukraine 252142

ABSTRACT. The concept of fundamental dynamic uncertainty (multivaluedness) developed in Parts I-III of this work and used to establish the consistent understanding of genuine chaos in Hamiltonian systems provides also causal description of the quantum measurement process. The modified Schrödinger formalism involving multivalued effective dynamical functions reveals the dynamic origin of quantum measurement indeterminacy as the intrinsic instability in the compound system of 'measured object' and (dissipative) 'instrument' with respect to splitting into spatially localised 'realisations'. As a result, the originally wide measured wave catastrophically (and really!) "shrinks" around a random accessible point thus losing all its 'nonlocal properties' with respect to other points/realisations. The dissipativity of one of the interacting objects (serving as 'instrument') is reduced to its (arbitrarily small) openness towards other systems (levels of complexity) and determines the difference between quantum measurement and quantum chaos, the latter corresponding to an effectively isolated system of interacting (micro-) objects. We do not use any assumptions on particular "classical", "macroscopic", "stochastic", etc. nature of the instrument or environment: physical reduction and indeterminacy dynamically appear already in interaction between two microscopic (quantum) deterministic systems, the object and the instrument, possessing just a few degrees of freedom a part of which, belonging to the instrument, should correspond to locally starting, arbitrarily weak excitation. This dynamically indeterminate wave reduction occurs in agreement with the postulates of the conventional quantum mechanics, including the rule of probabilities, which transforms them into consequences of the dynamic uncertainty.

NOTE ON NUMERATION OF ITEMS. We use the unified system of consecutive numbers for formulas, sections, and figures (but *not* for literature references) throughout the full work, Parts I-V. If a reference to an item is made outside its "home" part of the work, the Roman number of this home part is added to the consecutive number: 'eq. (12)' and 'eq. (12.I)' refer to the same, uniquely defined equation, but in the second case we know in addition that it can be found in Part I of the work.

*Address for correspondence: Post Box 115, Kiev - 30, Ukraine 252030.
 E-mail address: kiril@metfiz.freenet.kiev.ua

# 8. Incompleteness of quantum mechanics and the involvement of chaos

The ordinary scheme of quantum mechanics cannot provide non-zero complexity for Hamiltonian dynamics [1]; for any reasonable definition of complexity, this is equivalent to absolute dynamic predictability, reversibility, and the absence of the true chaos in quantum world (see the relevant discussions in sections 5.III and 6.III). The ensuing conflict with the predominantly chaotic behaviour of the counterpart classical systems (see e. g. [2-4]) leaves us with only two possibilities: either we have just an illusion of randomness and complexity in the form of classical chaos, while the world is basically predictable, or the standard quantum mechanics is not complete, at least for chaotic systems (see also section 1.I). At present, the choice within this alternative can be made rather by general considerations, largely non-physical and hardly rigorous. If one tends to accept that the highly inhomogeneous and self-developing world, filled with extremely sophisticated spatiotemporal mixture of organisation and irregularity, is more consistent with non-zero complexity, then he enters automatically into the *first level* of chaos involvement with the foundations of quantum mechanics. This position is well illustrated by the questions put in the title of an article on algorithmic complexity of quantum dynamics of classically chaotic non-dissipative systems (first paper in ref. [1]):

$$\text{Does quantum mechanics obey the correspondence principle ?} \\ \text{Is it complete ?} \qquad (35)$$

The definite (and negative) answers implied inevitably invoke, however, the third, hidden, question: is it the *same* incompleteness as that known before, existing from the birth of quantum mechanics and discovered by the Founding Fathers themselves? The rather positive answer of the authors of the cited paper can be certainly justified by the apparent awkwardness of the situation, where the fundamental incompleteness exists in several different types. However, as was shown in parts I-III above, one can propose certain natural modification of quantum formalism (represented by the Schrödinger equation) such that the correspondence principle is re-established in its conventional form also for chaotic systems. As this modified Schrödinger equation is obtained from its ordinary form by simple algebraic transformations and is reduced to it for regular dynamics, it inherits, a priori, all the basic axioms of the ordinary quantum mechanics including their incompleteness. At the same time we have no more contradictions, within this modified version, concerning the correspondence principle for dynamically complex systems. Therefore the 'chaos-induced' incompleteness can be removed, in a natural and self-consistent manner, providing in addition a variety of chaotic quantum dynamic regimes characterised by the well-defined physical complexity. In other words, the problem of quantum chaos can be resolved within the introduced concept of the fundamental dynamic uncertainty (or fundamental multivaluedness of dynamical functions, FMDF) which *does not depend* on the basic quantum postulates with all their strong and weak sides. These results permit one to advance in the positive direction while estimating the consistency of quantum mechanics (35):

$$\text{Quantum mechanics (in the modified form)} \\ \text{obeys the correspondence principle.} \quad (36) \\ \text{Still it is incomplete.}$$

From the other hand, it may seem that in this way the new 'chaotic' quantum mechanics delays possible solution for the old problem of incompleteness and even, in a sense, plays for the latter by permitting it to survive even the hard trial of quantum deterministic randomness. However, in this part IV we are going to show that, in fact, it is quantum-mechanical indeterminism, representing the most mysterious part of the fundamental quantum problems, that depends on, and can be effectively reduced to, the fundamental dynamical uncertainty revealed in parts I-III. In this way the *formal* incompleteness of quantum mechanics can be practically removed. This forms *the second level* of chaos involvement with the foundations of quantum mechanics.

We note at once that the general idea about the relation between the two types of uncertainty, the dynamic and the quantum ones, seems to be rather evident and has already been exploited, in both its directions (see e. g. [5]). However, these investigations only emphasised the conclusion following already from the general state of things in quantum chaos: one certainly could not reduce the true quantum indeterminacy[*)] to a basically regular regime of (conventional) quantum chaos (see e. g. [6-10]). Indeed, it is clear that the question itself about chaos involvement with the irreducible quantum indeterminacy can seriously be posed only when one has the *true* quantum chaos providing a fundamental dynamical source of randomness. This is exactly the case for the concept of FMDF presented in parts I-III, and in the next sections we propose the detailed realisation of the appearing hope to obtain a causal solution for the problem of quantum measurement. It is important that the price paid for this solution, the postulate of the fundamental dynamic uncertainty, seems to correspond well to the obtained results, in every sense: by its profound basic nature, universality, novelty, and practical efficiency discovered above.

This point of contact between wave mechanics and dynamic complexity could be anticipated not only from the side of deterministic randomness tending to find its place in quantum world, but also starting from intrinsic tendencies in

---

[*)] Note the difference between the fundamental property of unpredictability of detailed parameters of individual particle manifestation within the Schrödinger wave, which is referred to as *indeterminacy* in this paper, and the *uncertainty* of the observed wave characteristics entering Heisenberg's relations. Even if the two notions are not independent, the former concerns the fundamentals of the wave-particle dualism as they appear in the measurement process, while the latter concentrates more on the wave nature of quantum objects. In any case, within this paper we deal almost entirely with indeterminacy, and we prefer using this term to avoid any confusion, while leaving the term uncertainty for other related, but clearly distinct, concepts (e. g. the fundamental dynamic uncertainty of *any* chaotic dynamics which is supposed to be the *eventual* origin of quantum indeterminacy, cf. Part V).

understanding of basically irreducible roots of quantum indeterminacy. Indeed, after 70 years of quite intensive research on the problem, it persists in its paradoxical status of something that practically works perfectly well in an enormous variety of situations, but nevertheless cannot be understood at the level of the most general logic forming the base of the scientific method itself. The tensions created are so large that an appreciable part of quite serious approaches tends to admit the fundamental fail of this method, either by deducing that we deal with a basic cognition barrier (see e. g. [11] for the concept of the "veiled reality", *réel voilé*), or by replacing the escaping physical solutions with philosophic maxima (this was characteristic for the Bohr's position: "the task of physics is not to find how the nature is, but rather what can we say about it", see [12], p. 8), or even by linguistic arrangements (see e. g. [13]).

In this situation a practical agreement, the famous *Copenhagen interpretation*, has been accepted like a standard providing the optimal *compromise* between the known and the unknown that is involved with the as much famous *complementarity* principle of Bohr. To our opinion such understanding of this interpretation corresponds exactly to the role it really plays in quantum mechanics and permits one to avoid both typical extreme cases of confusion, where what is in fact just a reasonable (and temporary) compromise is misunderstood either as the ultimate fundamental truth, or as a gross blunder with tragic consequences. In general, after the long absence of a really self-consistent and fundamental enough solution for quantum mysteries, there is a tendency to treat this or that evidently partial solution rather as the final one that excludes further crucial progress.

It is not really surprising to find out that the existing exceptions from this rule lead to most creative approaches, though often underestimated and even disregarded, like that of de Broglie, even if neither of them has provided, up to now, the final answers. Indeed, it is well-known that a future physical concept takes its shape by properly formulated *questions* ensuing from the basically relevant, though often intuitive, *physical* thinking, rather than by immediate formal answers. This attitude, advocated so persistently by Louis de Broglie and starting always from the direct but subtle contact between Reason and Nature, has proven to be the only one capable of unravelling any great mystery. The solution to the problem of quantum indeterminism proposed below and its implications are also inspired by this *synthetic* approach, and we consider it to be not out of place to clearly designate this way of thinking as a general but indispensable base for the subsequent detailed constructions. It is quite natural that, as we shall see, the particular results obtained here are profoundly consistent with the causal wave mechanics of de Broglie, which leads finally to a tentative global scheme of the complete quantum theory, *quantum field mechanics* (section 10.V).

Independent of these our findings, it seems to be rather evident that the 70-year search for the complete quantum mechanics approaches now its turning point. One indication of it is felt from the character of internal development of the totality of numerous 'interpretations' whatever their relative popularity and conceptual weight within the whole field. Well illustrated by the recently appeared and differently oriented books [11,12,14][*], the current situation shows pronounced signs of basic saturation of principal point discussion both within each approach, and between them. From the other hand, the emerging new ideas of the 'physics of complexity' (see sections 1.I, 6.I) are sufficiently fundamental and revolutionary in order to ensure, after being presented in the proper form, a crucial positive advance of the most realistic causal interpretations. Before specifying the latter possibility in the next sections, we briefly outline some relevant details of a problem.

As was noticed by many researchers, the fundamental issues of quantum mechanics, and eventually the proposed postulates, fall naturally apart into two related but still distinct, and even opposed, components. This basic duality can be seen from different points of view, and the earliest and most general formulation is reduced simply to the fundamental wave-particle dualism. Indeed, if we take experimentally evident wave nature of micro-objects as an axiom, then the equally evident localised particle manifestations of the same objects are not only difficult to explain, but even taken as axiom they contradict the first one. While the practical-purpose standard interpretation gets out of it by taking, in fact, this contradiction as the main axiom, a non-contradictory solution to this basic problem is claimed to be found within the approach of D. Bohm proposed in 1952 and presented in the most complete form in [12]. In this interpretation the coexistence of waves and particles, both real, is postulated from the beginning (even though their detailed physical origin is *not* specified from the first principles, contrary to the complete version of de Broglie approach); of course, the two entities are coupled, and the proper choice of coupling seems to reconcile wave and particle aspects. However, the God appears to be more subtle than this. As various experiments show, not only particle and wave coexist, but in each its manifestation the particle 'kills' its wave without leaving any trace of it; the latter seems to be instantaneously shrinking around the particle showing itself in interaction with external objects: this is the famous 'wave reduction'. And as if it is not enough to bury any good theory, it appears that particle commits this unfaithful act in a provocatively irregular fashion: one can only know its probability depending on place and time, but never exactly where and when. Both these features cannot be intrinsically incorporated in the existing scheme of Bohm's causal interpretation: like so many other approaches it is finally forced to accept indeterminacy without fundamental physical substantiation, but rather in the form of a usual semiempirical probability definition, and to regard the reduction as a plausible consequence of omnipresent environmental influences [12]. The real, and positive, role of Bohmian mechanics is that it serves as the optimal, though not complete, reconciliation of a *mathematically* consistent description with the *physically* consistent causal *double-solution* interpretation of de Broglie conceived and further developed during practically all the period since the appearance of quantum mechanics (see e. g. [17-19]). We have here generally the same type of relation between Bohm's theory and the unreduced de Broglie approach as that between the standard Copenhagen interpretation and (complete) quantum mechanics in the whole, and it should be estimated with the same comprehension mentioned above (note that the so-called 'pilot-wave' interpretation, equivalent to the Bohmian mechanics, was proposed by de Broglie himself, 25 years before Bohm, as an explicitly simplified 'model' for the underlying complete version of the 'double solution', see e. g. e-print quant-ph/9902015).

---

[*] The excellent selections of other references on the subject can be found therein; recent review articles can be exemplified by refs. [15,16].

Thus the most mysterious among the observed quantum mechanical features are related to particle manifestations in physical interactions of its wave-pilot and comprise (wave) reduction and (particle emergence) uncertainty. These are the constituents of what is called quantum measurement, or rather the problem of measurement which resists to all attempts of consistent physical solution, as was explained above. Note that these difficulties do not directly touch the postulate about the wave implication and its dynamics as such. This leads us to another form of the above-mentioned duality of quantum problems appearing now as the couple 'wave dynamics - measurement process'. The first component in this couple does not generally invoke physical mysteries. In the usual scheme the wave dynamics is described by the Schrödinger equation (modified, according to the results of parts I-III, for chaotic systems). In the much more complete approach of de Broglie [17-19] it is the nonlinear wave dynamics which is supposed to give the *"double solution"*, composed of the quasi-linear Schrödinger $\psi$-function and the localised highly nonlinear soliton-like[*]) singularity ('particle'), these two parts forming a unique object. And although the precise mathematical description of this complex dynamics has never been found, these *technical* difficulties seem not to hide any *physical* puzzle. It is quite the contrary for the second dual component, the measurement process. The latter seems to be incomprehensible in terms of ordinary causal physics, and it is this contradiction which is at the origin of all the attempts to assume a 'particular quantum physics' and even a particular philosophy of quantum world. R. Penrose [14] emphasises the distinction between the paradoxes coming from quantum measurement, called **X**-mysteries, and unusual nonlocal (eventually, wave-related) quantum effects, **Z**-mysteries. In his remarkably precise general analysis he emphasises that the **Z**-mysteries can be understood, in principle, already within the existing physical and philosophical notions, though maybe not without considerable efforts, whereas the **X**-mysteries are basically incomprehensible within the current concepts, whatever the efforts, and need 'something else', some really new and fundamental notion-phenomenon-postulate(s), to be understood at the same level as other physical concepts.

---

[*]) We use the terms 'soliton-like' (solution) and 'quasi-soliton' for this peculiar part of the de Broglie double solution which was originally named "bunched wave" (*onde à bosse*), and later on referred to also as "singular wave" (*onde singulière*). We emphasize the difference of this object with respect to classical solitons known as exact solutions of certain nonlinear equations. In particular, quasi-soliton *must not* be an exact solution; it is rather an unstable (chaotic) solution of a nonlinear equation (see section 10.V). We still use the word 'soliton' in its designation referring to the property to form a high abrupt hump, though unstable.

In what follows we propose such a solution for the problem of quantum measurement (section 9). We shall see that it corresponds to the natural physical type of approach described above, being expressed, at the same time, in an unambiguous mathematical form. It is inevitably based on a new fundamental concept, that of the fundamental dynamic uncertainty, introduced in parts I-III and extended here to the case of quantum measurement. This solution does not directly concern the first part of the dual quantum problem, the wave postulate (though it basically permits of existence of a *material* wave), but we shall see that it is especially consistent with the causal physical picture of the double solution of de Broglie and effectively provides the real hope to specify and extend its formulation (section 10.I). In other words, we do try to 'cross off' the **X**-mysteries from the list of quantum mysteries, as it has been anticipated [14], to be left only with tractable (and now *easier* tractable), though non-trivial, **Z**-mysteries of nonlinear wave mechanics of real material field(s).

# 9. Chaos in quantum measurement: uncertainty and reduction

## 9.1. Theory of measurement beyond the standard interpretation and the fundamental multivaluedness of dynamical functions

Consider quantum system described by the wave function $\Phi(q)$, where $q$ varies in the configurational space corresponding to the representation of interest. We want to measure the quantity $f = f(q)$ for our system associated to the quantum-mechanical operator $\hat{f}$. The latter is characterised by the complete sets of its eigenfunctions $\{\phi_g(q)\}$ and eigenvalues $\{f_g\}$:

$$\hat{f}\phi_g(q) = f_g\phi_g(q) \ . \tag{37a}$$

Correspondingly, the wave function can be decomposed:

$$\Phi(q) = \sum_g a_g \phi_g(q) \ , \tag{37b}$$

where

$$a_g = \int \phi_g^*(q)\Phi(q)dq \ . \tag{37c}$$

Note that accepting these *mathematical* constructions confirmed by experiment, we do not insist, for the moment, on any particular physical interpretation for the wave function $\Phi(q)$. We just suppose that it *can* be given, eventually, some clear causal meaning, the same referring to the measured quantity $f$ and the associated operator. This eventual interpretation can be based, most probably, on the notion of the real material field (for $\Phi(q)$) and its measurable characteristics (like $|\Phi(q)|^2$), but we need not specify these details here. It will be seen that already the general analysis of the measurement process, within the proposed complex dynamics description, provides a transparent causal explanation for the most mysterious part of quantum behaviour, which in its turn essentially 'liberates' us for the unambiguous determination of the proper wave-function interpretation (see section 10.V).

We consider now what happens when our system, the *object* of measurement, enters in contact with the measurement *instrument*. By its level of generality this analysis is close to the standard consideration of quantum measurement (see e. g. [20], §7) as well as to that within many other interpretations (e. g. [12], chapter 8). However, we never take any 'comfortable', but inconsistently restrictive, assumptions about the instrument, the object of measurement, and their interaction; we try, instead, to *deduce* the specific properties of the measurement process as resulting basically from the (complex) dynamic behaviour of the combined quantum system. Neither we use the framework of any particular interpretation of quantum mechanics, though we do emphasize the consistency of our results with de Broglie approach. Our measurement instrument is nothing but another quantum system governed by the *same* deterministic laws as the object of measurement. It does not need to be "macroscopic" (as in reality the primary interactions, producing all the effect, always happen at the microscopic scale of the object), and what should be especially emphasised, the instrument is not at all "classical" in whatever special sense (even though it can be, as any quantum system, in a semiclassical regime, in the ordinary sense). In this way we avoid not only the evident vicious circle of the standard interpretation constrained of using the limiting case of quantum theory in the foundations of its general case, but also more elaborated schemes of randomness introduction 'by hand' in other interpretations.

Note also that we start with the abstract theory of measurement which is not formally confined to any particular measured quantity or experimental scheme. This will permit us to see better a number of general rules and, in particular, to establish a transparent relation with the Hamiltonian quantum chaos described in parts I-III. One should not forget, however, that according to the hypothesis of de Broglie [21] the fundamental quantum measurement always involves position as the measured quantity, which is intimately related to the double-solution picture. We specify our theory for this case with the help of a particular example of the position measurement in the next section 9.2, where the relation to the double-solution scheme is also discussed.

The instrument is characterised by a physical quantity $\Pi = \Pi(\xi)$ called *indication*, that depends on the instrument configuration $\xi$ and can be associated to the quantum-mechanical operator $\hat{\Pi}$. The latter possesses, quite similarly to the measured quantity $f$, its eigenfunctions $\{\psi_{nl}^0(\xi)\}$ and eigenvalues $\{\pi_n^0\}$:

$$\hat{\Pi}\psi_{nl}^0(\xi) = \pi_n^0 \psi_{nl}^0(\xi) \ . \tag{38}$$

The eigenvalues $\{\pi_n^0\}$ referred to also as *readings* can be thought of as (without actually assuming it at this stage) possible indications of the real instrument or, to be more precise, the values of the primary characteristics related eventually to actual indications. The existence of the second quantum number, $l$, enumerating the eigenfunctions means that readings are commonly degenerate: $l$ corresponds to the internal instrument-related degrees of freedom which are not directly registered in detail, but are necessary for the physical functioning of the instrument. For example, in a position-measuring device $l$ may enumerate different states of the detector atoms, whereas $n$ corresponds to the positions of these atoms (see the next section). It is clear that $l$ should take at least two values designating the unexcited and excited states of the instrument elements.

The plurality of readings expresses the unreduced quantum nature of the instrument. However, in the absence of object any real properly "tuned" instrument should almost certainly find itself in a particular state, the "ground state" $\psi_{in}^0(\xi)$, well separated from all other states:

$$\psi_{in}^0(\xi) = \frac{1}{\sqrt{N_\Pi}} \sum_n \psi_{n0}^0(\xi) \ , \tag{39a}$$

where $N_\Pi$ is the number of possible indication values ensuring the wave function normalisation (it should normally, but not necessarily, be equal to the number, $N_f$, of the measured eigenvalues), and the unexcited state of the instrument element corresponds to $l = 0$. This initial instrument state $\psi_{in}^0(\xi)$ is represented typically by the real ground state of all atoms of the detector, etc., so that without the exciting action of the measured object (e. g. a projectile) the instrument can remain in its ground state practically forever without revealing its quantum nature (apart from small noises, etc.). So the wave function of the whole system *before* the measurement is:

$$\Psi_0(q,\xi) = \Phi(q)\psi_{in}^0(\xi) , \tag{39b}$$

where $\Phi(q)$ can be presented in the form (37b).

Now the process of measurement is determined by the *interaction* of the measured object with the instrument, specified by its operator $V$ depending on $f$, $\Pi$ and acting in the mixed configurational space $q \otimes \xi$, $V = V_f(q,\xi)$. This process can be described both in time-dependent and time-independent versions of quantum formalism, which are effectively equivalent.[*] Here we limit ourselves with the latter as it seems to be more suitable for our purposes. The starting equation is obtained as a natural supposition that the combined system of object interacting with instrument is characterised by some time-independent eigenstate $\Psi(q,\xi)$:

$$[f(q) + \Pi(\xi) + V_f(q,\xi)]\Psi(q,\xi) = \pi\Psi(q,\xi) . \tag{40}$$

It is at this point and in this way that we fix the fact of the given measurement with its particular instrument, quantum object, and measured quantity for this object. Note that in many cases the operators $f(q)$ and $\Pi(\xi)$ can be reduced to the Hamiltonians of the respective systems, while $V_f(q,\xi)$ represents the effective potential of their interaction, and then eq. (40) is simply the stationary Schrödinger equation for the whole system. In its general form eq. (40) encompasses also possible deviations from this standard situation.[#]

Further analysis resembles much the method of the effective dynamical functions used to reveal Hamiltonian quantum chaos (see section 2.I-II). We shall see now that the same line of arguments leads to similar conclusions for the measurement process. We first expand the total wave function in the complete set of functions $\{\phi_g(q)\}$ associated to the measured quantity $f$:

---

[*] In particlar, this was demonstrated for the case of Hamiltonian quantum chaos in parts I-II.

[#] One may consider also that eq. (40) corresponds, in fact, to the description of a dissipative (open) system already reduced to the non-dissipative one in order to include explicitly only the essential degrees of freedom of the instrument (i. e. without the explicit analysis of the basically important excitation processes, see the discussion of the nondemolition measurement below in this section). This reduction is physically reasonable and can be done by the well-known methods (e. g. by the standard effective potential method, see [22]).

$$\Psi(q,\xi) = \sum_g \psi_g(\xi)\phi_g(q) , \qquad (41)$$

where $\psi_g(\xi)$ are just decomposition coefficients to be determined from eq. (40). Substituting expansion (41) into eq. (40), multiplying the result by $\phi_g^*(q)$, and integrating over $q$, we obtain the system of equations for $\{\psi_g(\xi)\}$:

$$[\Pi(\xi) + V_{00}(\xi)]\psi_0(\xi) + \sum_g V_{0g}(\xi)\psi_g(\xi) = \pi_0 \psi_0(\xi) , \qquad (42a)$$

$$[\Pi(\xi) + V_{gg}(\xi)]\psi_g(\xi) + \sum_{g' \neq g} V_{gg'}(\xi)\psi_{g'}(\xi) = \pi_g \psi_g(\xi) - V_{g0}(\xi)\psi_0(\xi) , \qquad (42b)$$

where

$$\pi_0 \equiv \pi - f_0 , \quad \pi_g \equiv \pi - f_g , \qquad (43)$$

$$V_{gg'}(\xi) \equiv \int_{\Omega_q} dq\, \phi_g^*(q) V_f(q,\xi) \phi_{g'}(q) , \qquad (44)$$

$\Omega_q$ is the domain of a function under the integral, and we have separated the case $g = 0$, considering from now on that $g, g' \neq 0$ in all equations, by definition.

We try now to reduce the system (42) to equation only for $\psi_0(\xi)$ using the familiar substitution method (see section 2.1.I). The Green function for the homogeneous part of eq. (42b) is

$$G_g(\xi,\xi') = \sum_n \frac{\psi_{gn}^0(\xi)\psi_{gn}^{0*}(\xi')}{\pi_{gn}^0 - \pi_g} , \qquad (45)$$

where $\{\psi_{gn}^0(\xi)\}$ and $\{\pi_{gn}^0\}$ are the complete sets of eigenfunctions and eigenvalues for the auxiliary system of equations,

$$[\Pi(\xi) + V_{gg}(\xi)]\psi_g(\xi) + \sum_{g' \neq g} V_{gg'}(\xi)\psi_{g'}(\xi) = \pi_g \psi_g(\xi) . \qquad (46)$$

(Note that here and below we can omit quantum number $l$ including it into $n$, for simplicity.)

Using the well-known properties of the Green function, we express $\psi_g(\xi)$ through $\psi_0(\xi)$ from eqs. (42b):

$$\psi_g(\xi) = -\int_{\Omega_\xi} d\xi'\, G_g(\xi,\xi') V_{g0}(\xi') \psi_0(\xi') , \qquad (47)$$

where $\Omega_\xi$ is the domain of a function under the integral. Substituting this expression into eq. (42a), we obtain the desired *effective measurement equation* for $\psi_0(\xi)$:

$$[\Pi(\xi) + V_{\text{eff}}(\xi)]\psi_0(\xi) = \pi\psi_0(\xi) , \qquad (48)$$

where the *effective interaction* instrument-object, $V_{\text{eff}}(\xi)$, is given by

$$V_{\text{eff}}(\xi) = V_{00}(\xi) + \hat{V}(\xi) , \quad \hat{V}(\xi)\psi_0(\xi) = \int_{\Omega_\xi} d\xi' V(\xi,\xi')\psi_0(\xi') , \qquad (49a)$$

$$V(\xi,\xi') \equiv \sum_{g,n} \frac{V_{0g}(\xi)\psi_{gn}^0(\xi)V_{g0}(\xi')\psi_{gn}^{0*}(\xi')}{\pi - \pi_{gn}^0 - f_g} . \qquad (49b)$$

(Note that eqs. (48), (49) are written with the assumption that $f_0 = 0$ which can always be satisfied by the agreement about the eigenvalue numbering, and the latter is often naturally gives that result. In any case, the assumption is made for simplicity and cannot influence the conclusions.)

Equation (48) provides the complete sets of solutions, $\{\pi_n\}$ and $\{\psi_{0n}(\xi)\}$, which can be used then to obtain the general solution of a problem with the help of expansion (41). We express first other wave-function components $\psi_g(\xi)$ through the $\psi_{0n}(\xi)$ found, using eq. (47):

$$\psi_{gn}(\xi) = \hat{\eta}_{gn}\psi_{0n}(\xi) \equiv \int_{\Omega_\xi} d\xi' \eta_{gn}(\xi,\xi')\psi_{0n}(\xi') , \qquad (50a)$$

$$\eta_{gn}(\xi,\xi') = \sum_{n'} \frac{\psi_{gn'}^0(\xi)V_{g0}(\xi')\psi_{gn'}^{0*}(\xi')}{\pi_n - \pi_{gn'}^0 - f_g} . \qquad (50b)$$

We can construct now the general solution of the initial measurement problem, eq. (40):

$$\Psi(q,\xi) = \sum_n c_n \Big[\phi_0(q) + \sum_g \phi_g(q)\hat{\eta}_{gn}\Big]\psi_{0n}(\xi) . \qquad (51)$$

The coefficients $c_n$ here should be found from the boundary conditions which consist in the wave-function matching along the boundary determined by the condition $V_f(q,\xi) = 0$. Finally, the experimentally observed quantity should be compared to $\rho(q,\xi) \equiv |\Psi(q,\xi)|^2$. Of course, for the measurement process considered one of the most important observable quantities is provided by the instrument readings $\{\pi_n\}$.

We see thus that the properties of the solution (51) are determined eventually by those of the effective interaction $V_{\text{eff}}(\xi)$ and the corresponding effective measurement equation, eqs. (48), (49). As it was demonstrated in Part I for the case of Hamiltonian quantum chaos (see section 2.2.I), the specific feature of the effective functions of this type is that they are generally splitted into many 'independent' and 'incoherent' components, together with solutions of the equations involved. We are going now to re-establish this *fundamental dynamic uncertainty* (or the fundamental multivaluedness of dynamical functions, FMDF) in terms of our quantum measurement problem. The main idea is that in the modified formulation, eqs. (48)-(51), of the initial problem, eq. (40), the eigenvalues to be determined, here $\pi$, enter also the expression for the effective function, eq. (49), in the form of nonlinear multi-branch dependence, $V(\xi,\xi') \equiv V(\xi,\xi';\pi)$. This leads to self-consistent splitting, $V_{\text{eff}}(\xi) \to \{V^i_{\text{eff}}(\xi)\}$, $\{\pi_n\} \to \{\pi^i_n\}$, $\{\psi_{0n}(\xi)\} \to \{\psi^i_{0n}(\xi)\}$, where the components, $\Re_i \equiv \{V^i_{\text{eff}}(\xi), \{\psi_{0n}(\xi)\}^i, \{\pi_n\}^i\}$, numbered by the superscript $i$, contain the complete, in the ordinary sense, sets of solutions and are called *realisations*. Then the expectation value of experimentally observed quantity should be represented as a sum over the realisations (for more detail see eq. (16.II) and the accompanying discussion, section 2.3.II):

$$\rho(q,\xi) = \sum_i \alpha_i \rho_i(q,\xi) \equiv \sum_i \alpha_i |\Psi_i(q,\xi)|^2 ,$$

$$\Psi_i(q,\xi) = \sum_n c^i_n \left[ \phi_0(q) + \sum_g \phi_g(q) \hat{\eta}^i_{gn} \right] \psi^i_{0n}(\xi) , \quad \sum_i \alpha_i = 1 , \tag{52}$$

where $\alpha_i$ is the probability of occurrence of the $i$-th realisation, and $\hat{\eta}^i_{gn}$ is obtained from $\hat{\eta}_{gn}$, eqs. (50), by substitution of $\pi^i_n$ for $\pi_n$. Of course, one can speak about the expectation value only for *many* repetitions of the experiment for the same conditions. For each *individual* experimental run one will observe *only one* of the branches $\{\rho_i(q,\xi)\}$ (because each of them is a complete solution, in the ordinary sense) appearing *at random* (because none of them is preferable in any sense) with the respective probabilities $\{\alpha_i\}$. This basically provides already a solution to the problem of origin of quantum-mechanical indeterminacy which is reduced, in this way, to the fundamental dynamic uncertainty. Before developing further arguments in favour of this assertion, we briefly describe the demonstration of the reality of the underlying FMDF, quite similar to that for the Hamiltonian quantum chaos (section 2.2.II).

We start with rewriting eq. (48) for the $n$-th eigenfunction, $\psi_{0n}(\xi)$, multiplying it by $\psi_{0n}^{*}(\xi)$, and integrating over $\xi$, which gives:

$$\mathcal{V}_{nn}(\pi_n) = \pi_n - \Pi_n^0 , \qquad (53)$$

where

$$\mathcal{V}_{nn}(\pi_n) \equiv \sum_{g,n'} \frac{|V_{nn'}^g|^2}{\pi_n - \pi_{gn'}^0 - f_g} , \qquad (54)$$

$$V_{nn'}^g \equiv \int_{\Omega_\xi} d\xi\, \psi_{0n}^{*}(\xi) V_{0g}(\xi) \psi_{gn'}^0(\xi) , \qquad (55)$$

and

$$\Pi_n^0 \equiv \int_{\Omega_\xi} d\xi\, \psi_{0n}^{*}(\xi)[\Pi(\xi) + V_{00}(\xi)] \psi_{0n}(\xi) . \qquad (56)$$

The obtained form, eq. (53), of the effective measurement equation, eq. (48), is especially suitable for the graphical analysis. We plot, in Fig. 2, the left- and right-hand sides of eq. (53) as functions of $\pi_n$, the abscissas of the intersection points of the two curves giving the real eigenvalues for the instrument readings.[*] It is clear that the number of these solutions is determined by the number of branches of the function $\mathcal{V}_{nn}(\pi_n)$, eq. (54). Namely, if $N_f$ and $N_\Pi$ are the numbers of terms in the sums over $g$ and $n'$, respectively, in eq. (54), then the number of solutions is

$$N_s = N_f N_\Pi + 1 . \qquad (57a)$$

From the other hand, the maximum 'normal' number of solutions for the initial system of equations (42) is

$$N_s^0 = N_\Pi + 1 \qquad (57b)$$

(it is sufficient to compare eqs. (42) with the auxiliary system, eqs. (46), providing $N_\Pi$ terms in the sum over $n'$ in eq. (54)). The difference $\Delta N_s \equiv N_s - N_s^0 = N_\Pi(N_f - 1)$ cannot be less than 1, as we suppose that the measured quantity $f$ possesses more than one eigenvalue, $N_f \geq 2$, and the instrument is capable to provide at least one indication, $N_\Pi \geq 1$. In reality one has practically always $N_\Pi > 1$ and often $N_\Pi, N_f \gg 1$, whereupon $\Delta N_s \gg N_s^0 \gg 1$.

---

[*] Note that the values of $\Pi_n^0$ can differ somewhat for different $\pi_n$, which means that the segments of the line $\pi_n - \Pi_n^0$ between the neighbouring asymptotes will be slightly vertically displaced one relative to another. It is easily seen, however, that this modification does not influence the results obtained, and we do not show it in Fig. 2 to avoid unnecessary complications.

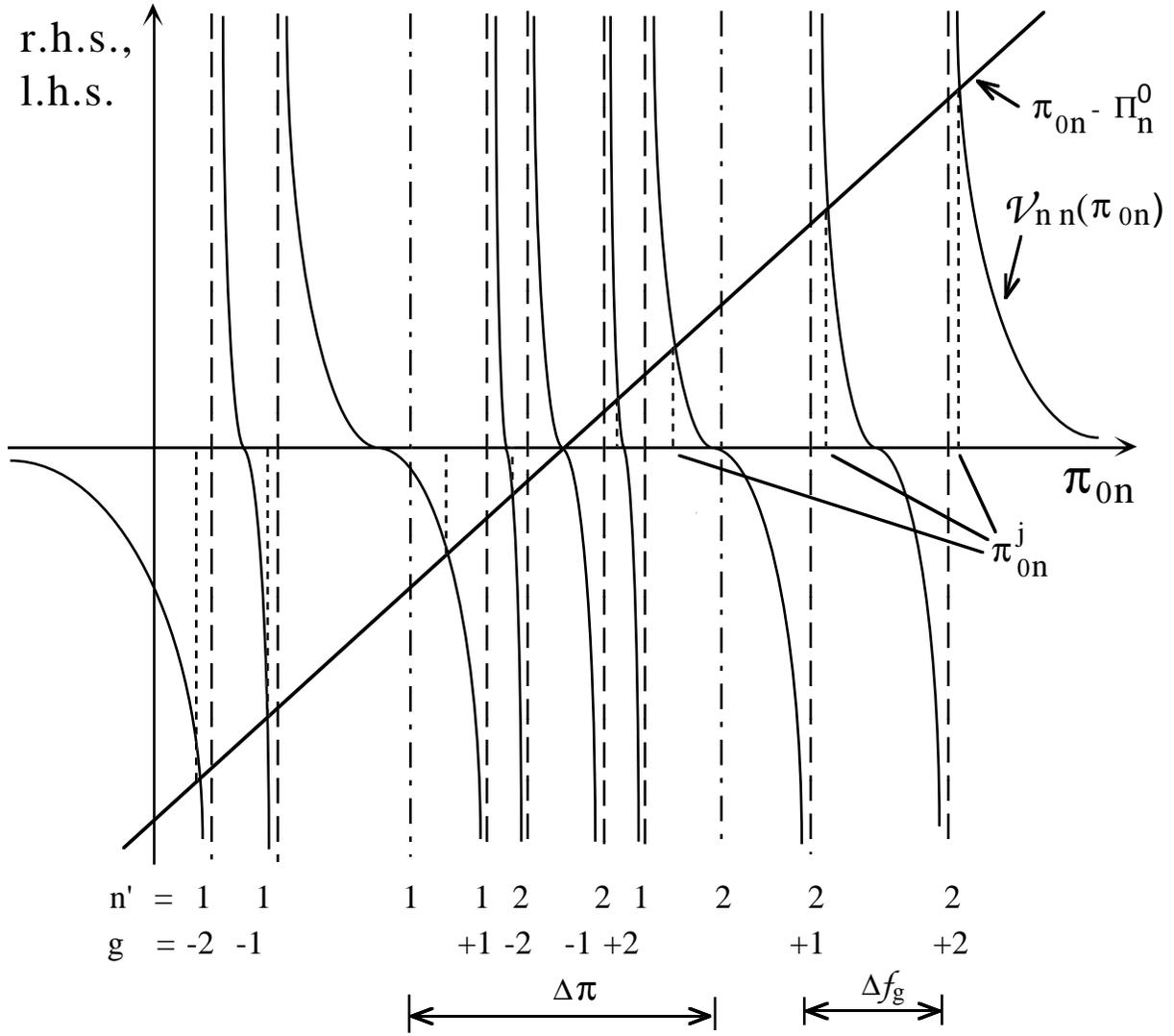

**Fig. 2**. Scheme of the graphical solution of the effective measurement equation, eq. (53).

Moreover, eqs. (57) show that one has rather multiplication, than addition, of solutions while passing from ordinary to the effective measurement equations. This corresponds to the above idea about the multiple *realisations* of the system, each of them consisting of the normal complete set of $N_s^0$ solutions. The number of realisations is

$$N_\Re = N_s/N_s^0 \approx N_f .\qquad(57c)$$

This grouping of solutions has basically important physical meaning that leads to the causal interpretation of the 'mystery' of quantum measurement.

Indeed, as can be seen from eqs. (53)-(54), using also the definitions (37), (41), and (48), certain *i*-th realisation $\Re_i$ ($1 < i < N_\Re$), comprising the complete set of $N_s^0$ solutions with its instrument readings, $\{\pi_n^i\}$, is related to one of the eigenvalues, $f_g$, of the measured quantity, $f$, because these readings are obtained from the branches of $\mathcal{V}_{nn}(\pi_n)$ for this fixed $g$ and varying $n'$ in eq. (53) (see Fig. 2). In other words, there is a one-to-one correspondence between the realisations of the combined system object-instrument (numbered by $i$) and the eigenvalues of the measured quantity (numbered by $g$): $\mathfrak{R}_i \leftrightarrow f_g$, $i \equiv i(g)$, $g \equiv g(i)$. The different indications $\pi_n^i$ within each group ($i$ is fixed, $1 < n < N_\Pi$) correspond to usually insignificant and practically often not distinguished instrument-related variations, such as different excitation energies transmitted by the object to the electrons of the same instrument atom, or different, but close, excited atoms of the same detector.

It is extremely important, and it strictly follows from the formalism above, that each of the realisations is formed by the *complete* set of eigenfunctions, so that *only one* of them can be realised in the individual experimental run. (One may say also that the realisations are mutually *incoherent*.) As a priori all the realisations are equally probable, this gives us the result exactly identical to quantum indeterminacy as it appears in the real acts of measurement. Namely, at each individual experimental run one of the $N_\mathfrak{R}$ complete sets of the combined-system eigenfunctions and eigenvalues (indications), corresponding to one of the $N_f$ eigenvalues of the measured quantity $f$ in the measured state $\Phi(q)$, is realised *randomly* (i. e. one can know only the probability of that event), which is naturally fixed as the unpredictable (though expected) appearance of *this* eigenvalue. Note that in terms of our general quantum chaos theory (see section 3.II), this situation of quantum measurement automatically falls in a region just below the 'classical' border of global chaos, $\Delta\pi \sim \Delta f$ ($\Delta\pi$ and $\Delta f$ are the respective eigenvalue separations), where the randomisation is at its maximum. This relation between $\Delta\pi$ and $\Delta f$ follows directly from the definition of the instrument function, and the inference obtained explains why the indeterminacy of quantum measurement is basically irreducible.

The relation between the realisations and the measured eigenvalues, revealed above, is definitely manifested also at the level of eigenfunctions. As is seen from eqs. (49), (50), when a realisation number, $i$, is fixed, the operators $\hat{V}^i(\xi)$, $\hat{\eta}_{gn}^i$, and thus the wave-function components $\psi_{gn}^i(\xi)$, attain their maximum magnitude for $g = g(i)$ and certain $n = n(i)$ and diminish from this maximum with growing differences $|g - g(i)|$ and $|n - n(i)|$. Then it is clear from eqs. (52) that the occurrence of the $i$-th realisation in an experimental run will be accompanied by a partial *reduction* (absolutely causal!) of the total wave function (51) to a small number of its components centred around

$$\Psi_{ign}(q,\xi) = \phi_g(q) \, c_n^i \hat{\eta}_{gn}^i \psi_{0n}^i(\xi) \,, \quad g = g(i) \,, \quad n = n(i) \,, \tag{58a}$$

or rather to

$$\rho_{ig}(q,\xi) = |\Psi_{ig}(q,\xi)|^2 \equiv |\sum_n \Psi_{ign}(q,\xi)|^2 \,,$$

because the realisations are incoherent, they correspond to observable quantities. The (random) manifestation of the value $f_g$ ($g = g(i)$) of the measured quantity is thus evident both from the eigenvalue and eigenfunction behaviour of the solutions of the measurement problem in its modified formulation, eqs. (48)-(52). This causal quantum indeterminacy is inseparably linked to the accompanying wave-function reduction, eq. (58a).#) Note, however, that the reduction obtained in this way from the modified Schrödinger formalism is only partial, which means that the total wave function is reduced to a group of *several* eigenfunctions, giving a *group* of eigenvalues as the result of measurement, rather than to the unique respective quantities determined by the strict condition $g = g(i)$. As can be seen from eqs. (48), (50), (58a) with the background of Fig. 2, in the generic case the observed quantity components for the $i$-th realisation, $\rho_{ig}(q,\xi)$, diminish faster than $(\Delta g)^2/[g - g(i)]^2$ around the maximum at $g = g(i)$, where $\Delta g \sim 1$ is the neighbouring eigenvalue separation. Properly conducted experiments could help deciding whether the obtained reduction is sufficient to explain the observed behaviour of quantum systems.*)

Even apart from this eventual experimental verification, one can see another reason for the wave-function reduction based on the combination of the presented formalism with the de Broglie idea about the existence of a nonlinear soliton-like singular part of the wave which is represented by the small region of very high intensity surrounded by a relatively smooth distribution of the low-intensity Schrödinger wave. Indeed, this singular wave-field structure should correspond to a very narrow range of $g$ values or even to a single value.°) Then, according to eq. (44), only a few matrix elements $V_{g0}(\xi)$ will be different from zero, 'activating' the corresponding small number of the wave function components $\psi_{gn}^{i}(\xi)$, eqs. (50). It is important to note that this physically transparent reduction mechanism, always implied in the de Broglie approach, cannot be really successfully applied without the formalism of the fundamental dynamical uncertainty. Indeed, our solution, eqs. (52), for the modified measurement equation provides the real, causal physical reduction of the wave function avoiding the problem of the remaining coherent "empty wave" typical for the causal dualistic wave-particle theories (see e. g. [12]). We continue to clarify this relation with the double-solution concept below in terms of particular measurement processes (section 9.2) and within our synthetic scheme of quantum field mechanics (section 10.V).

The revealed causal reduction involves also the configurational behaviour of the instrument wave functions. It can be specified if one assumes that the

---

#) Note that in the realistic case of position measurement the same phenomena can be described in terms of the effective nonlinear wave instability (see section 9.2 for more details).

*) One should certainly take into account that in most real measurement processes the density of the measured eigenvalues is rather high (already because the instrument should normally be quasi-classically large).

°) It seems to be quite probable that in this fashion de Broglie singularity determines, in fact, the hidden very fine-scale discretion of all quantities superimposed on the ordinary continuous and discrete spectra.

effective configurations of the instrument are of the same type as the measured quantity (it is normally true for the most important case of coordinate measurement). This implies, referring to eq. (39a), that the initial instrument function, $\psi_{\text{in}}^0(\xi)$, is a superposition of a number of *localised and well-separated* (with respect to $\xi$) components, $\psi_{n0}^0(\xi)$, the property that guaranties the necessary *resolution* of the instrument. Normally the configuration of instrument includes a number of identical sensitive units corresponding to different indications, and then $\psi_{n0}^0(\xi) = \psi_0(\xi - \xi_n)$, where $\psi_0(\xi)$ is a δ-like function with the effective width $\Delta\xi \ll |\xi_n - \xi_{n+1}|$.[#] This property is evidently inherited by the interaction potential $V_f(q,\xi)$ and thus by the effective measurement equation solutions, $\{\psi_{0n}^i(\xi)\}$, as it can be seen from eqs. (44), (48), (49). Namely, eq. (49b) shows that $\psi_{0n}^i(\xi)$ is localised around certain $\xi = \xi_g$, where $g = g(i)$, $n = n(i)$. Thus the 'eigenvalue' reduction of the total wave function to a small number of its components, centred around a randomly chosen eigenvalue, is accompanied by its physical squeeze around the point of localisation of those components.

Now what has remained to do for completion of our causal description of indeterminacy and reduction is to deduce the quantum-mechanical (Born's) rule of probabilities of the measured eigenvalue appearance. We have discovered above that the fundamental dynamic uncertainty provides the causal basis for the random manifestation of different eigenvalues, $f_g$, of the measured quantity in the form of incoherent localised realisations. The realisations appear with a priori equal probabilities. However, in practice they are distributed with a high and inhomogeneous density, so that experimentally one can observe only groups of close realisations and not the individual ones (this situation is discussed also in section 2.3.II, within the general formulation of FMDF). Each group has a fixed size in configurational space, but contains varying number of realisations. Then the real probabilities, $\alpha_i$, refer to such groups of realisations and can evidently be calculated by counting the number of individual realisations within each group. As we are going to show, in our current problem this can be automatically achieved by satisfying the condition of wave-function matching at the boundary between the instrument and the 'free' object. Using eqs. (58a), (37), (39), we derive this condition in the form that accounts for the reduction described above:

$$a_g \psi_{\text{in}}^0(\xi_0) = \sum_{n \approx n(i)} c_n^i \hat{\eta}_{gn}^i \psi_{0n}^i(\xi_0) , \quad g = g(i) , \qquad (59)$$

where $\xi_0$ is in general a function of $q$ determining the boundary surface, and the sum at the right-hand side includes, in fact, the summation over that smallest experimentally resolvable interval of realisations which corresponds to the same discernible value of $g$.

---

[#] In this case the possible instrument indications, $\pi_n^0$ (see eq. (38)), can evidently be identified with the respective localisation points $\xi_n$.

Recall now that the functions of $\xi_0$ at both sides of eq. (59) are the sums of highly localised components, and the detailed dependence on $\xi_0$ within each component is rarely of interest. To exclude this dependence we integrate eq. (59) over an interval of $\xi_0$ large enough and centred around $\xi = \xi_g$ ($g = g(i)$) renormalising, if necessary, the constants $c_n^i$:

$$a_g = \sum_{n \approx n(i)} c_n^i \equiv C_i , \quad g = g(i) . \tag{60}$$

It is not difficult to see that because of the hidden effective summation over the irresolvable individual realisations mentioned above, the quantities $C_i$ thus defined should be in fact replaced, in the expressions for the observed quantity, eqs. (52), with the probabilities of realisation occurrence,

$$\alpha_i = |C_i|^2 = |a_g|^2, \tag{61}$$

while

$$c_n^i = C_i c_n , \quad \sum_{n \approx n(i)} c_n = 1 ,$$

where the coefficients $c_n$ characterise the details of configurational dependence within localised (groups of) realisations (they may only slightly depend on $i$, as the registration elements of the instrument are identical).

The conventional quantum-mechanical rule of probabilities is thus *causally explained* within our approach: the probabilistic character of quantum behaviour is due to the fundamental dynamic uncertainty of the excessive choice of realisations, and the probability of occurrence of the $g$-th eigenvalue ($\equiv$ occurrence of the $i$-th group of realisations, $g = g(i)$) is equal to $|a_g|^2$, for the measured wave function of eq. (37b). In particular, the Born's normalisation rule is an evident consequence of the general realisation probability normalisation, eq. (16b.II). It is not difficult to see that in this way we arrive at a basic *physical* definition of the notion of probability itself referring to a universal definition, and a method of derivation, of elementary event ($\equiv$ realisation) as a causally deduced dynamical object. This provides a rigorous basis for theoretical probability calculation without making reference to experimentally determined possible outcomes in the system behaviour or to any empirical rule including Born's rule and the law of large numbers (it is a general property of the fundamental dynamic uncertainty, see sections 2.3.II and 6.III).

The local details of the wave-function dependence on $\xi$ can be specified by multiplying the boundary condition, eq. (59), by $\psi_{0n}^{i*}(\xi_0)$ and integrating it over the boundary surface. We can find then the values of $c_n$ which transform eq. (58a) into

$$\Psi_{ig}(q,\xi) = \phi_g(q) \sum_{n \approx n(i)} \frac{\hat{\eta}_{gn}^i \psi_{0n}^i(\xi) \int\limits_{\xi=\xi_0} d\xi\, \psi_{0n}^{i*}(\xi_0) \psi_{\text{in}}^0(\xi_0)}{\int\limits_{\xi=\xi_0} d\xi\, \psi_{0n}^{i*}(\xi_0) \hat{\eta}_{gn}^i \psi_{0n}^i(\xi_0)} , \quad g = g(i). \qquad (58b)$$

We have seen above that for a large class of measurements $\psi_{0n}^i(\xi)$ is localised around certain $\xi = \xi_g$, where $g = g(i)$, $n = n(i)$. Then eq. (58b) shows how this $\xi_g$ 'matches' to the corresponding $\xi_n$ ( $\equiv \pi_n^0$) due to the cutting integration in the numerator. Thus the above boundary condition application provides *both formal and physical* correspondence between the $n$-th sensitive unit (possible indication) of the instrument and the $g$-th measured eigenvalue represented by the reduced wave function $\Psi_{ig}(q,\xi)$, where $g = g(i)$.

The presented formalism can be used also for clarification of a number of subtle aspects of quantum measurement related, in particular, to the so-called 'nondemolition' measurements. We start with noticing that we could perform our analysis of the initial measurement equation (40) by expanding the total wave function similarly to eq. (41), but using the complete set of functions $\{\psi_n^0(\xi)\}$ as the basis, instead of $\{\phi_g(q)\}$. This corresponds, formally, to the interchange of the variables q and $\xi$, so that we could repeat all the stages of the method of effective dynamical functions to arrive finally at expressions analogous to eqs. (48)-(52), but with the interchanged $q$ and $\xi$.

Physically, however, this interchange is much less symmetrical. Indeed, the instrument, contrary to the object, contains certain *internal* (local) degrees of freedom that can be *excited* towards a configurational manifold *open to the outside*, the property which makes the instrument an *open (dissipative) system* accessible to the *observer* (see below) and characterised by the effective *internal (local) irreversibility* provided the external degrees are *actually* cut from *detailed* observation. This property should not be confused with the *global dynamic irreversibility* of a system originating from the fundamental dynamic uncertainty. One may suppose that the irreducible local dissipativity of the instrument can eventually be provided with more fundamental physical grounds, appealing to the interaction of the de Broglie soliton-like 'humps', within the physically complete description of quantum field mechanics (section 10.V).

In the case corresponding to starting expansion in the form of eq. (41) (we shall call it *real measurement*) one obtains, in physical terms, a complete set of incoherent realisations of the *instrument* (variable $\xi$), each of them appearing randomly and related, as it was shown above, to its 'preferred' eigenvalue (and eigenfunction) of the quantity of interest for the object (variable $q$) which is thus really 'measured'. In the other case corresponding to starting expansion over $\{\psi_n^0(\xi)\}$ (we call it *nondemolition* or *fictitious measurement*) one obtains a complete set of incoherent realisations of the *object* (variable $q$), each appearing randomly and related to a particular state $\psi_n^0(\xi)$, $\pi_n^0$ of the *unexcited* instrument, and not to an eigenvalue of some quantity for the object. In this second case the instrument does not experience any *local* irreversible change of the dissipative type, contrary to the first situation, even though its unexcited state after a fictitious measurement may still unpredictably differ from that before the measurement. In return, the fictitiously measured object undergoes an irreversible change of its *global* state by 'falling' into one, or even many, successive random and incoherent realisations, so that something 'serious' (irreversible) does happen to the object, but it is not unambiguously registered by the instrument (the latter should be irreversibly *excited* to provide any distinct registration). In addition, such nondemolition global 'reduction' of the object wave function will not be accompanied by localisation of the reduced wave function characteristic for the real measurement. This is because in the case of fictitious measurement there is no any symmetry-breaking localised internal excitations, and the random occurrence of a realisation is accompanied by a delocalised change in the wave-function structure described by our theory of quantum chaos for non-dissipative systems (Parts I-II, see also the discussion below). The term (non)demolition refers thus both to the instrument and the object meaning the (non)excitation of particular localised degrees of freedom of the instrument and the (non)localisation of the object wave function. Note that the irreversibly changed state of the object after the nondemolition measurement can be identified with the help of another, 'demolition' (real) measurement, which confirms that fictitious measurement is a quite real, and measurable, physical process. It is not clear to which degree our definition of nondemolition measurement can be compared to the existing meanings of the term (see e. g. [23]), rather ambiguous in themselves, but our definition seems at least to be consistent by referring to the proposed causally complete concept of quantum measurement.

One may inquire, however, about a peculiar situation, where a change of expansion basis in the mathematical formalism leads to physically different consequences. The answer is that, for each particular real situation, we are not actually free to choose one or another form of the formalism. Namely, if some internal irreversible physical changes do happen to the instrument during its interaction with the object then the instrument can no more be described by the unperturbed states, eq. (38), and the corresponding eigenfunctions cannot be used as a basis in the starting wave-function expansion. Physically, it means that although some wave functions can still be associated with the instrument, it may only be the *effective* eigenstates of the *combined* system object-instrument, eq. (48), falling apart, as we have seen, into the incoherent localised realisations due to the instability of excessive choice. If the instrument remains 'untouched' internally, both expansions can formally be used, but the version oriented for the real measurement, eq. (41), may eventually provide only one realisation for the instrument because of the particular, 'separable' form of the effective interaction $V_f(q,\xi)$ (see section 9.2). In this case it is more pertinent to use the expansion over the eigenfunctions of the unperturbed instrument, eq.

(38), leading, in the general case, to the 'non-dissipative' quantum chaos (Parts I-II).

One can express this difference between the real and fictitious measurements in another way by noting that the real measurement, accompanied by irreversible internal changes in the instrument, corresponds to the irreducible modification of the configurational variable $\xi$ itself: $\xi \rightarrow \xi \otimes \xi'$, where $\xi'$ characterises the excited degrees of freedom (the latter are, however, non-stochastical). It is thus equivalent to the irreducible involvement of excitation processes, whatever weak, or in other words, to the fact that the instrument is basically an open system. It is clear that in this case the expansion on any complete set of functions depending only on $\xi$ will not be efficient because the coefficients will depend on both $q$ and $\xi'$.

The condition for the instrument to be an open system has a number of profound implications. First of all, in the opposite case the instrument could not be used as such. The latter implies, however, the indirect introduction of the *observer* in our scheme, just as an open part of the instrument that *cannot* be described by the *same* wave function as the instrument. This reminds us about the well-known 'anti-causal' interpretations of quantum mechanics essentially involving the subjective human-based influences (e. g. of consciousness) on the measurement process. In this connection we should emphasize the basic difference of our approach with respect to those interpretations: in our description the existence of the observer is strictly limited to the irreducible property of the instrument to be an open system, with or without the actual conscious being interacting with it, whereas the fundamental randomness, and thus the global irreversibility, of the measurement process is due to the dynamic FMDF mechanism specified above. From the other hand, such indirect 'causal' implication of observer in quantum measurement seems to be both unavoidable and useful, for many reasons. In particular, it provides the universality to our description which does not depend on the magnitude of the excitation processes serving as an issue from the instrument to the outside; we demand only the existence of this issue, whatever narrow it may be.

In general, it is important to note that both presented interpretations of the physical difference between the real and nondemolition measurements satisfy naturally the peculiar limitation on any theory of quantum measurement: the fact of measurement as such, with the accompanying reduction and indeterminacy, should not depend on the magnitude of interaction between object and instrument including the limit of very small magnitudes. This specific feature could not be consistently interpreted within the existing approaches, either causal or formal. In our approach, even the smallest *real* (i. e. excitation-like ≈ locally irreversible ≈ "open") interaction induces fundamental changes both in the proper description choice and in the global character of the occurring physical processes. This peculiarity may have further physical implications within de Broglie conception of the singular soliton-like 'particle' (section 10.V).

These notions of real and nondemolition measurements, ensuing from our concept of the fundamental dynamic uncertainty, permit one to clarify also the relation between dynamical quantum chaos in general and its fundamental manifestations in the measurement processes. We see now that all the basic physical processes involve chaotic wave dynamics. However, one can objectively distinguish two manifestations of quantum chaos. One of them corresponds to the above nondemolition (fictitious) measurement and is in fact the 'ordinary' (but true!) quantum chaos of the non-dissipative, Hamiltonian type (described in Parts I-II) observed in the behaviour of an object interacting with some 'instrument' that does not actually serve as an instrument in the sense of quantum measurement; it remains effectively 'inert' (non-dissipative). In the other situation 'something' internally (and locally) irreversible really happens to the instrument during its interaction with the object, and although it results basically in the same phenomenon of global dynamical instability by the FMDF mechanism, it appears now in the form of real measurement with reduction and 'quantum' indeterminacy (which is just a manifestation of the fundamental dynamic uncertainty). One may say that quantum measurement is quantum chaos in open systems with dissipation which differs from the 'pure' quantum chaos in Hamiltonian systems by localisation during system reduction to a (randomly chosen) realisation.

To avoid any misunderstanding, note that in the situation of real measurement, the fundamental dynamical randomness is not reduced at all to the effect of the irreversible excitations of the instrument (already because they can be infinitesimal), even though they induce the important difference between the two cases of quantum dynamic complexity. We may also point out that there can always exist a subjective (and rather trivial) practical aspect of the relation between the two cases: one actually measures what he wants to measure, and for example in the majority of real physical experiments there are lots of acts of quantum measurement which are simply not registered as such.

A mixture of the two manifestations of quantum chaos can also exist: one can perform quantum measurement over the quantum chaotic object. In this case one can imagine that the role of the regular object above is played by one of the realisations of the chaotic object, so that one will measure the characteristics of this particular realisation, and there will be double (quantum) randomness in the results of such measurement (if the realisations of the chaotic object and of the combined system are more entangled, then it is simply not the best choice of the instrument).

It is worthwhile to note that this implication of chaos in the behaviour of elementary quantum objects, in its both manifestations described above, can certainly be employed to specify the notion of quantum-mechanical transition and its probability (for example, those for tunnelling, see the end of section 4.III); this is one of the subjects for further investigation within the 'chaotic' quantum mechanics mentioned at the end of section 7.III.

Note finally that the specific property of the instrument (or rather that of the whole measurement process) which effectively replaces the demand for the instrument to be "classical", or "macroscopic", within the existing theories of measurement (e. g. [20,24]), is its 'chaoticity'. The latter is not related to any additional limitation on the instrument, but is just a consequence of the fact that the whole system <object + instrument> possesses, practically always, more than one degree of freedom. Typically the total number of the degrees of freedom is even much larger than one, but we do not demand this in our description. In reality, the act of quantum measurement as such involves essentially a limited number of 'primary' degrees of freedom (in principle, it can be the electronic configurations of just one atom, or of other *elementary quantum* object serving as a measurement instrument, - see section 9.2). The other degrees of freedom, participating in further amplification and registration of a signal within the measurement set, do not directly influence quantum indeterminacy and wave reduction (that is why they could be effectively eliminated from the above analysis, eq. (40)), though being quite indispensable practically and basically (to maintain the open character of the whole system). What plays effectively the role of multiple and randomising 'environmental' degrees of freedom, necessarily referred to in other measurement theories, is the multiplicity of our realisations which can be 'occupied' by the system only at random. This intrinsic chaoticity is causally deduced within our modified formalism through the effective measurement equation, eq. (48), leading automatically to the dynamic indeterminacy of the localisation point of the reduced wave. The latter phenomenon has practically unlimited universality and corresponds exactly to the semiempirical rules of the standard interpretation.

## 9.2. Example of measurement with intrinsic uncertainty and reduction: particle coordinate measurement

Consider an example of a simple quantum system, where the process of measurement can be especially transparent, in order to illustrate the above general results by their specific manifestations within a particular physical picture. One of the most popular demonstrations of quantum-mechanical concepts involves a projectile wave-particle incident on the opaque obstacle with two open holes (or slits); the wave will then penetrate the holes and form an *interference pattern* in the distribution of the point-like *registration events* (visible for a large number of them) on a sensitive screen placed at a certain distance behind the obstacle. We introduce just a little complication in this scheme partially closing each hole with a particle-sensitive detector (Fig. 3). The material and sensitivity of the detectors are so chosen that there is an appreciable probability for a particle to traverse a hole without any excitation of the detector atoms; the reverse is also quite probable, and then the detector that has been excited register the passage of a particle and maybe even some of its characteristics (e. g. energy).

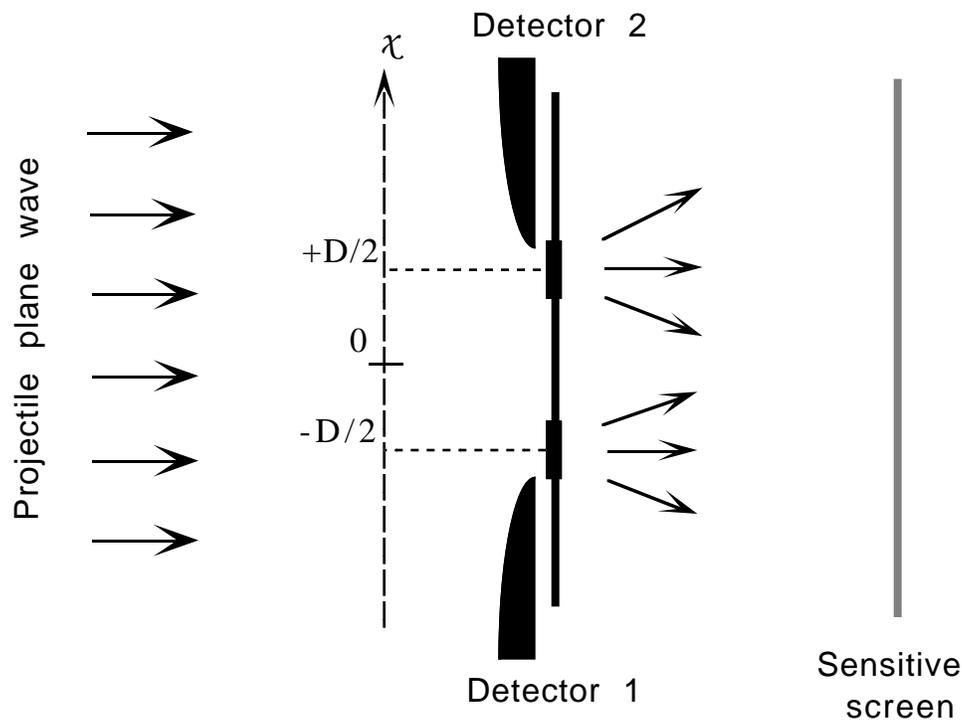

**Fig. 3**. Classical two-slit experiment slightly modified for the particle coordinate measurement.

According to the rules of quantum mechanics, based on many different experimental verifications, there are only the following two possibilities for the behaviour of this system: either the wave-particle passes freely through the holes producing a point-like registration event on the sensitive screen, the density of those events from many such freely passed particles being distributed according to the interference pattern for wave diffraction on a system of the two holes; or else there is the change of state of one of the detectors indicating the passage of the particle, and then the other detector will never be activated, for the same wave-particle, and the density of registration events of many such particles inducing the excitation of one of the detectors, will be quasi-homogeneous, without any trace of diffraction on the two holes. For a passage of a single particle, one can never know exactly whether one of the two detectors will come into action and which one, nor at which point of the screen the registration event will happen; in return, the probabilities of each of these events can be known, in principle, exactly.

The disappearance of the interference pattern on the screen in the second case (one of the detectors comes into action) cannot be explained by perturbation of the wave diffraction process during its interaction with the detector. Indeed, it is well known that, for example, for energetic particles, or waves, the relative perturbation during scattering, $\delta$, is of the order of $U/E$ ($U$ is the effective interaction potential, $E$ is the particle energy), whether it concerns amplitudes, phases, trajectories, or angles. From the other hand, the characteristic size, $\Lambda$, of the interference-pattern inhomogeneities is proportional to $1/D\sqrt{mE}$, where $D$ is the distance between the holes, and $m$ is the mass of the particle. It is then clear that one can always choose the parameters, such as $U$, $E$, $D$, and $m$, in a range where $\delta$ is much smaller than $\Lambda$.

This means that there is indeed a basic contradiction between the two possible types of behaviour described above, free-wave diffraction and measured-particle transmission. It is this peculiarity that is referred to as the wave-particle dualism, and is expressed also as the irreducible and mysterious reduction and indeterminacy. The standard, or formal, interpretations of quantum mechanics just take them for granted. But neither can they be fully understood within the existing causal schemes of quantum mechanics. In one of the interpretations of de Broglie double solution (see [17-19]) the distributed Schrödinger wave is considered rather as a sort of fiction, mathematically useful, but physically not necessary existing. One may inquire, however, what is the exact relation between this fictitious wave obeying the Schrödinger equation and physically real soliton-like singular wave structure, the second part of the double solution. One possibility is that this soliton-like 'particle' is in a state of permanent quasi-random rapid motion remaining unresolved ('hidden') within the experimentally observed smooth Schrödinger wave. The invoked source of this spontaneous motion, the "hidden thermodynamics", necessitates the existence of the "hidden thermostat", another postulated entity which corresponds to a yet more fundamental level of being, the "subquantum medium". But what is especially disturbing is that this "fictitious" smoothing wave produces really observed physical diffraction effects.

It is probably one of the circumstances which motivated David Bohm to postulate, within his version of the same idea of the causally dual object, the existence of both distributed Schrödinger wave and point-like particle, without specifying their profound physical nature. However, even apart from the arising questions about this nature, the formally dualistic solution of the ensuing "quantum theory of motion" [12] cannot describe neither reduction, nor indeterminacy without again making reference to omnipresent stochastic environmental influences (another "hidden thermostat"!) and imposing a number of assumptions and arbitrary parameters (see section 8 in [12]). In our general theory of measurement naturally following from the universal concept of the fundamental dynamical uncertainty we have shown above (section 9.1) how one can avoid the contradictions of indeterminacy and reduction remaining in accord with experimental results, as well as with the principle of causality. Now we are going to provide a particular physical illustration of these general results taking the simple quantum system considered (Fig. 3) as an example.

The measurement by one of the detectors placed in the holes determines, when it takes place, the position of the particle within the incident wave, $\Phi(z) = \Phi_0(x)\exp(ikz)$, where the coordinate axis $z$ is perpendicular to the plane, $(x,y)$, of the obstacle with the $x$ axis passing through both holes, $\Phi_0$ is some known function ($\Phi_0(x) \equiv$ const for the case of the incident plane wave), and we can omit the dependence on $y$ due to the geometry of a problem. In this case $q \equiv x$ in our general description of the measurement process, section 9.1. The measured quantity is the coordinate $x$, and its operator, $\hat{x}$, is characterised by the following eigenfunctions and eigenvalues in the coordinate representation (cf. eq. (37a)):

$$\hat{x}\,\delta(x - x´) = x´\delta(x - x´) . \tag{62a}$$

Expansion (37b) is reduced to the δ-function definition:

$$\Phi_0(x) = \int_{-\infty}^{\infty} dx´\,\Phi_0(x´)\delta(x - x´) , \tag{62b}$$

meaning simply that the particle is situated somewhere within the interval $-\infty < x < \infty$ with the probability density distribution

$$a_x = \Phi_0(x) . \tag{62c}$$

Note that we have chosen here $z = 0$ for the position of the plane of the obstacle. We shall generally exclude the coordinates $y$ and $z$ from our consideration, where their role is insignificant for the measurement process analysis.

The indication operator $\hat{\Pi}$ for the detectors is also reduced to the operators of coordinates, those of the electrons of detector atoms (cf. eq. (38)):

$$\hat{X}\,\delta(X - X_{nn´}) = X_{nn´}\delta(X - X_{nn´}) , \tag{63}$$

where $X_{nn´} \equiv X_n + \xi_{n´}$, $X_n$ is the position of the centre of the n-th detector, and $\xi_{n´}$ is the position of the $n´$-th electron of each detector with respect to its centre.

The wave function of the whole system before the measurement is evidently (cf. eq. (39))

$$\Psi_0(x,X) = \Phi(x,z)\overset{0}{\psi}_{\text{in}}(X) ,$$

where (64)

$$\overset{0}{\psi}_{\text{in}}(X) = \sum_{n,n'} \overset{0}{c}_{nn'}\delta(X - \overset{0}{X}_{nn'}) , \quad \overset{0}{X}_{nn'} \equiv X_n + \overset{0}{\xi}_{n'} ,$$

$\overset{0}{\xi}_{n'}$ runs through all possible electron positions in the unexcited detector, and $\overset{0}{c}_{nn'}$ are the (known) numerical components of the wave function decomposition.

The interaction between the projectile and the detectors is described by certain operator $V(x,X)$ with maximum magnitude at $x \cong X$ that decreases rapidly with growing $|x - X|$. To obtain a consistent description one should suppose also that this operator is nonlocal in $x$:

$$V(x,X)\Psi(x,X) \equiv \int_{-\infty}^{\infty} dx'\, \vartheta(x,x',X)\Psi(x',X) .$$

We need not specify here any particular form for the kernel, $\vartheta(x,x',X)$. This nonlocality can be formally reduced to the Heisenberg uncertainty manifestation for the measured wave function: any real interaction, involving local excitation processes (see section 9.1), necessitates fixing the measured projectile momentum, and this is equivalent to delocalisation of the corresponding effective interaction (see the footnote after eq. (40)). It is quite probable, however, that at the more fundamental level this is related to the specific properties of the nonlinear double solution in de Broglie's picture, and in this case it is one of the connection points between the non-local quasi-linear Schrödinger description and the anticipated nonlinear local description (see the corresponding discussion below, section 10.V). Another fundamental origin of nonlocality of the interaction operator $V(x,X)$ is also due to the reduced character of the starting measurement equation, eq. (40), where the degrees of freedom of the local excitations are effectively replaced by the relevant properties of the interaction operator; the corresponding nonlocality has a short-range character, contrary to the nonlocal effects of the double-solution. Note that whatever its origin the nonlocality itself does not introduce any reduction or additional indeterminacy. In return, it emphasises the fundamental role of the local excitation processes transforming the fictitious measurement (non-dissipative quantum chaos) into the real measurement (section 9.1). Indeed, in the absence of those processes the interaction operator, $V(x,X)$, loses its nonlocality in $x$, and the measurement problem becomes separable (in the system (68) below $V_{gg'}(X) = V_{gg}(X)\delta(g - g')$, and the equations become independent).

The starting measurement equation for the wave function of the combined system projectile-detectors is (cf. eq. (40)):

$$[\hat{x} + \hat{X} + V(x,X)]\Psi(x,X) = \chi\Psi(x,X) . \tag{65}$$

Following our general method, we expand $\Psi(x,X)$ in the complete set of functions $\{\delta(x - x')\}$:

$$\Psi(x,X) = \int_{-\infty}^{\infty} dx' \psi_{x'}(X)\delta(x - x') = \psi_x(X) . \tag{66}$$

We can now considerably simplify the formulae, while preserving all the essential details, if we suppose that the components $\psi_x(X) = \Psi(x,X)$ have appreciable magnitude only for $x \approx X_1$ and $x \approx X_2$, i. e. within each of the holes-detectors. As we are not interested in the detailed wave-function structure within the detectors, we may use the 'point-hole' approximation considering that $\psi_x(X)$ has only three components, $\psi_1(X)$, $\psi_2(X)$ and $\psi_0(X)$. The last component, $\psi_0(X)$, corresponds to the part of the wave function which 'does not traverse the holes', i. e. to eigenvalues $x = X_0 \neq X_1, X_2$ (without it the model would be too simple to provide the unambiguous manifestation of the dynamic uncertainty). This means that $g = 0, 1, 2$ in our general formalism above ($g = 1, 2$ in most expressions), and eq. (66) can be rewritten as (cf. eq. (41)):

$$\Psi(x,X) = \psi_0(X)\phi_0(x - X_0) + \psi_1(X)\phi_1(x - X_1) + \psi_2(X)\phi_2(x - X_2) , \tag{67}$$

where $\phi_g(x - X_g)$ are the characteristic functions essentially differing from zero within the corresponding intervals centred around $x = X_g$. Physically $\psi_g(X)$ describes the wave-particle passing through the $g$-th detector (where the 0-th detector is just the impenetrable obstacle) or, to be more precise, the probability density of that event, and *a priori all those g possibilities are quite coherent and superposable*, as follows directly from eqs. (66), (67). Note also that this elementary model with the three-component wave function, eq. (67), is especially useful for the transparent demonstration of the causal dynamic mechanism of reduction and indeterminacy; from the other hand, it will automatically provide evidence for the universality and really basic character of the fundamental dynamic uncertainty which permits the same solution of the measurement problem even in such a simple, a priori deterministic system with very few degrees of freedom. As concerns the extension of the results to a more realistic description, it is rather straightforward and can be achieved by specifying the general expressions of section 9.1 to a particular model (see also the end of this section).

After this simplification the system of equations (42) for the wave-function components is reduced to three equations:

$$[\hat{X} + V_{00}(X)]\psi_0(X) + V_{01}(X)\psi_1(X) + V_{02}(X)\psi_2(X) = \chi_0\psi_0(X) , \tag{68a}$$

$$[\hat{X} + V_{11}(X)]\psi_1(X) = \chi_1\psi_1(X) - V_{10}(X)\psi_0(X) , \tag{68b}$$

$$[\hat{X} + V_{22}(X)]\psi_2(X) = \chi_2\psi_2(X) - V_{20}(X)\psi_0(X) , \tag{68c}$$

where

$$V_{gg'}(X) = \int_{-\infty}^{\infty}\int_{-\infty}^{\infty} dx\,dx'\, \phi_g^*(x - X_g)\vartheta(x,x',X)\phi_{g'}^*(x' - X_{g'}) \cong L_x \vartheta(X_g, X_{g'}, X), \qquad (69)$$

$L_x$ is a normalisation constant, and

$$\chi_g \equiv \chi - X_g. \qquad (70)$$

The general character of the function $V_{gg'}(X)$ is determined by its extrema (normally minima) at $X = X_g, X_{g'}$. One can choose the origin of $x$ so that $X_0 = 0$, $X_1 = -D/2$, $X_2 = +D/2$, where $D$ is the distance between the detectors; then $\chi_0 = \chi$, $\chi_1 = \chi + D/2$, $\chi_2 = \chi - D/2$. In order to simplify technical details we consider the case when $|V_{12}(X)| \ll |V_{10}(X)|$, which is a natural assumption taking into account the origins of the nonlocality described above. This facilitates the reduction of system (68) to one equation, but does not influence any essential physics (all the results are reproduced also without this simplification, in agreement with the general scheme, section 9.1).

In accord with our general procedure, we use now the method of substitution with the help of the Green functions for the homogeneous parts of eqs. (68b,c) which are:

$$G_g^0(X,X') = \sum_n \frac{\psi_{gn}^0(X)\psi_{gn}^{0*}(X')}{\chi_{gn}^0 - \chi_g}, \qquad (71)$$

where $\{\psi_{gn}^0(X)\}$, $\{\chi_{gn}^0\}$ are the eigen-solutions for

$$[\hat{X} + V_{gg}(X)]\psi_g(X) = \chi_g \psi_g(X). \qquad (72)$$

This equation is just a little bit more complicated than the initial equation for the unperturbed detectors, eq. (63). It is the rough "mean-field approximation" for the detectors interacting with a projectile, this interaction being averaged over the unperturbed states of the projectile, $\phi_g(x) \approx \delta(x - X_g)$. Therefore we can suppose that the solutions $\{\psi_{gn}^0(X)\}$, $\{\chi_{gn}^0\}$ can be relatively easily found and that they are close to those for the unperturbed detectors (cf. eq. (64)):

$$\psi_{gn}^0(X) \approx \phi_{gn}^0(X - X_n) \approx \delta(X - X_n), \quad \chi_{gn}^0 = X_n + \Delta_{gn}, \quad |\Delta_{gn}| \ll |X_n|, \quad n,g = 1,2. \qquad (73)$$

So we express $\psi_{1,2}(X)$ through $\psi_0(X)$ from eq. (68b,c):

$$\psi_{1,2}(X) = -\int_{\Omega_X} dX'\, G_{1,2}^0(X,X') V_{10,20}(X')\psi_0(X'). \qquad (74)$$

Substituting it into eq. (68a) we obtain the desired effective measurement equation for $\psi_0(X)$:

$$[\hat{X} + V_{\text{eff}}(X)]\psi_0(X) = \chi\psi_0(X) , \qquad (75)$$

where the effective interaction projectile-detectors, $V_{\text{eff}}(X)$, is deduced as

$$V_{\text{eff}}(X) = V_{00}(X) + \hat{V}(X) , \quad \hat{V}(X)\psi_0(X) = \int_{\Omega_X} dX' V(X,X')\psi_0(X') , \qquad (76)$$

$$V(X,X') \equiv \sum_{n=1}^{2} \left( \frac{V_{01}(X)\psi_{1n}^{0}(X)V_{10}(X')\psi_{1n}^{0*}(X')}{\chi - \chi_{1n}^{0} + D/2} + \frac{V_{02}(X)\psi_{2n}^{0}(X)V_{20}(X')\psi_{2n}^{0*}(X')}{\chi - \chi_{2n}^{0} - D/2} \right).$$

To complete the solution, we find the eigenfunctions, $\{\psi_{0n}(X)\}$, and the eigenvalues, $\{\chi_n\}$, of eq. (75) and we use them to obtain $\psi_{1,2}(X)$ with the help of eq. (74):

$$\psi_{1,2n}(X) = \hat{\eta}_{1,2n}(X)\psi_{0n}(X) \equiv \int_{\Omega_X} dX' \eta_{1,2n}(X,X')\psi_{0n}(X') , \qquad (77a)$$

$$\eta_{1,2n}(X,X') = \sum_{n'=1}^{2} \frac{\psi_{1,2n'}^{0}(X)V_{10,20}(X')\psi_{1,2n'}^{0*}(X')}{\chi_n - \chi_{1,2n'}^{0} \pm D/2} . \qquad (77b)$$

We can finally write down the general solution:

$$\Psi(x,X) = \sum_n c_n \left[ \phi_0(x - X_0) + \delta(x - X_1)\hat{\eta}_{1n}(X) + \delta(x - X_2)\hat{\eta}_{2n}(X) \right]\psi_{0n}(X). \qquad (78)$$

To reveal the fundamental multivaluedness we reproduce the 'graphical representation', eqs. (53)-(56):

$$\mathcal{V}_{nn}(\chi_n) = \chi_n - X_n^0 , \qquad (79)$$

where

$$\mathcal{V}_{nn}(\chi_n) \equiv \sum_{n'=1}^{2} \left( \frac{|V_{nn'}^{1}|^2}{\chi - \chi_{1n'}^{0} + D/2} + \frac{|V_{nn'}^{2}|^2}{\chi - \chi_{2n'}^{0} - D/2} \right), \qquad (80)$$

$$V_{nn'}^{1,2} \equiv \int_{\Omega_X} dX\, \psi_{0n}^{*}(X) V_{01,02}(X) \psi_{1,2n'}^{0}(X), \tag{81}$$

and

$$X_n^0 \equiv \int_{\Omega_X} dX\, \psi_{0n}^{*}(X)[\hat{X} + V_{00}(X)] \psi_{0n}(X). \tag{82}$$

To specify our graphical analysis recall that each of the sets $\{\chi_{1n'}^0\}$, $\{\chi_{2n'}^0\}$ contains the corresponding two eigenvalues, in accord with eqs. (73). This leads to the graphical representation of eq. (79) shown in Fig. 4.[*] We find five solutions for $\chi_n$, denoted as $\chi_n^i$, which correspond to two different realisations (enumerated by the superscript $i$), according to the general formulae, eqs. (57). Each realisation is composed from the normal complete set of three eigenvalues and eigenfunctions. The eigenvalues $\chi_1^1$, $\chi_2^1$ represent the first realisation, and the eigenvalues $\chi_1^2$, $\chi_2^2$ correspond to the second one. The fifth solution for $\chi_n$, the one close to zero, belongs to both realisations and does not play a crucial role by itself; for the strictly symmetrical zero-order approximation, when $\Delta_{gn} \equiv 0$ in eqs. (73), this solution disappears. As it is implied by eqs. (65), (68), the quantities $\chi_n \equiv \chi_{0n}$ provide physically possible values of instrument indications $X$ with respect to $X = X_0 = 0$, i. e. in fact the resulting values of the measured projectile coordinate $x$. It is then easy to see that the first realisation ($i = 1$) incorporates the solutions $\{\chi_1^1, \chi_2^1\}$ related to the first detector (the asymptotes around $\chi_n \sim x = -D/2$ in Fig. 4), and the second realisation is similarly related to the second detector. As the realisations are independent and incoherent (because each of them is complete), we may interpret this *eigenvalue evidence* as the fact that the wave-particle within the $i$-th realisation *passes* through the $i$-th detector.

One can now specify the more profound origin of such excessive choice of solutions, giving rise to the independent incoherent realisations, for this particular system. It is related to the fact that the positions of our detectors (i. e. the observed measurement indications) enter *twice* into the measurement equation in the modified, effective form (see eq. (75)), first as the detector positions themselves and second as *possible* projectile positions during measurement. The irreducible inelastic interaction between the object and the instrument, whatever its magnitude, effectively 'entangles' these five coordinate eigenvalues from the double 'redundant' set, so that they all should appear in the resulting observation which cannot contain more than three of them, the complete set for our model. This expresses exactly the real physical choice that one *would* have for a localised *particle* (for example, for that of the de Broglie double solution) interacting with the two holes-detectors. The modified formulation of a problem provides the corresponding two *independent incoherent* realisations starting from the Schrödinger *coherent* (and initially linear) *wave* description and using causal (and rather simple) considerations. The agreement thus obtained between the two dualistic approaches inherent to quantum mechanics, that of the linear non-localised wave and that of the non-linear localised particle, is not a coincidence; it leads to the unified scheme of 'quantum field mechanics' incorporating the two pictures, the local and the nonlocal ones, as its dual constituents and thus resolving the basic mystery of the wave-particle dualism (see section 10).

---

[*] Note that as the value of $X_n^0$, eq. (82), may vary for different solutions $\chi_n^i$, the segments of the line $\chi_n - X_n^0$ intersecting the respective branches of the function $\mathcal{V}_{nn}(\chi_n)$ in Fig. 4 can be slightly displaced vertically one relative to another. However, this will not produce any significant changes on the scale of our schematical representation, neither in the conclusions obtained, and we do not show these secondary details to avoid unnecessary complication.

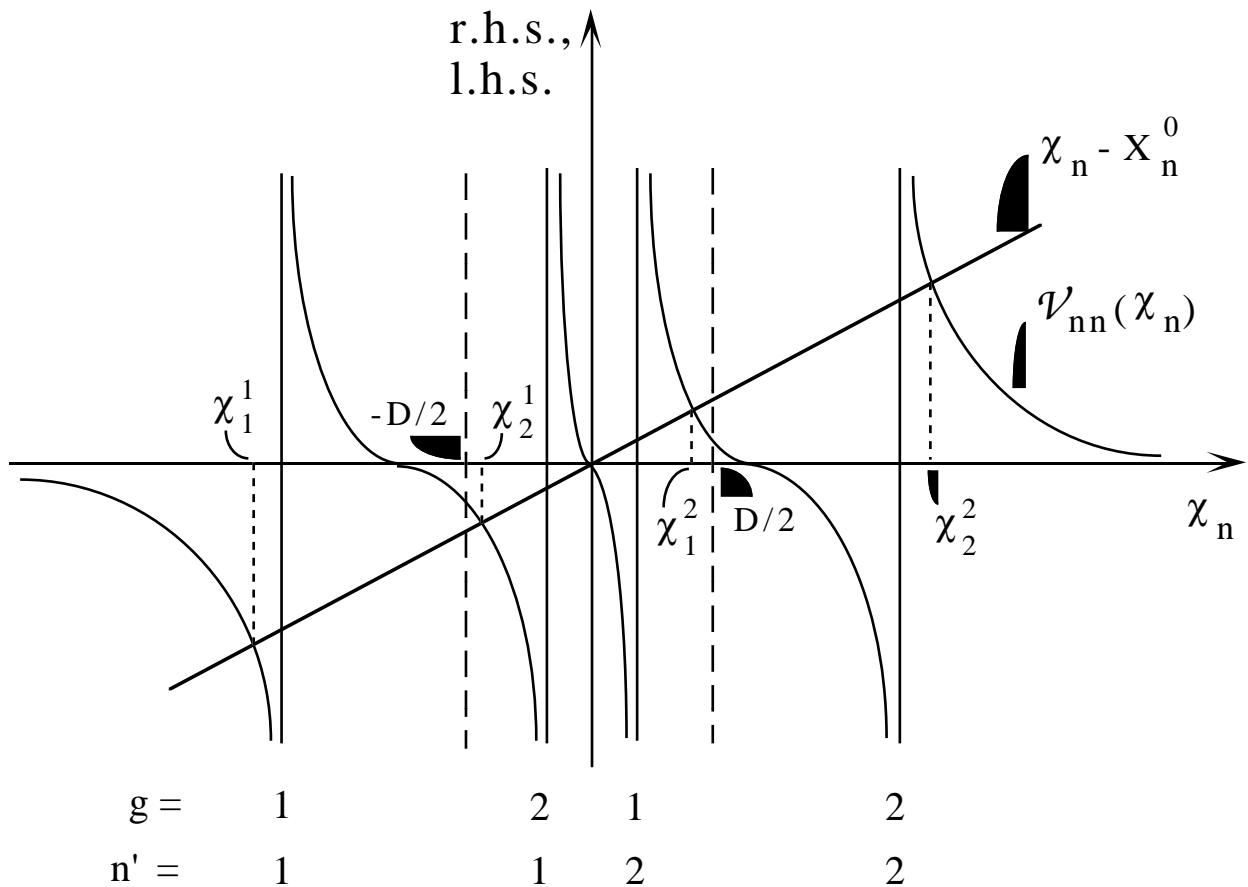

**Fig. 4**. Graphical solution of the modified measurement equation, eq. (79), for the coordinate measurement in the two-slit experiment, Fig. 3. We plot the right- and the left-hand sides of eq. (79) vs the eigenvalue, $\chi_n$, to be determined.

The discovered fundamental multivaluedness in the measurement process permits us to write down the total solution for the observed wave-field intensity (eqs. (52) in the general formalism):

$$\rho(x,X) = \rho_1(x,X) \oplus \rho_2(x,X), \quad \rho_{1,2}(x,X) \equiv |\Psi_{1,2}(x,X)|^2, \qquad (83)$$

$$\Psi_i(x,X) = \sum_{n=1}^{2} c_n^i \left[ \phi_0(x - X_0) + \delta(x - X_1)\hat{\eta}_{1n}^i(X) + \delta(x - X_2)\hat{\eta}_{2n}^i(X) \right] \psi_{0n}^i(X) ,$$

where $\oplus$ stands for the probabilistic sum, $i = 1, 2$, and the kernels of the operators $\hat{\eta}_{gn}^i(X)$ are

$$\eta_{1,2n}^i(X,X') = \sum_{n'=1}^{2} \frac{\psi_{1,2n'}^0(X) V_{10,20}(X') \psi_{1,2n'}^{0*}(X')}{\chi_n^i - \chi_{1,2n'}^0 \pm D/2} . \qquad (84)$$

The *causally* obtained *probabilistic* sum in eqs. (83) certifies that the measured passage of the projectile through one or another detector occurs really at random, though with certain well-defined probabilities $\{\alpha_1, \alpha_2\}$ that can be determined (see below). Now to arrive at a completely self-consistent picture, we should provide a *wave-function evidence* for the passage of the wave-particle in the $i$-th realisation through the $i$-th detector. This evidence is especially simple for the considered coordinate measurement and consists in the causal wave localisation accompanied by the dynamic indeterminacy in the position of the centre of this localisation.

Indeed, as follows directly from eq. (8O), for the $i$-th realisation the $i$-th term in the sum over $g$ in the expression for the wave function, eq. (83), is much greater than the others, and so the total wave function is well approximated by this term:

$$\Psi_i(x,X) \approx \Psi_i^{\text{red}}(x,X) = \sum_{n=1}^{2} c_n^i \delta(x - X_i) \hat{\eta}_{in}^i(X) \psi_{0n}^i(X) , \qquad (85a)$$

while the observed intensity is

$$\rho_i(x,X) \approx \rho_i^{\text{red}}(x,X) = |\Psi_i^{\text{red}}(x,X)|^2 . \qquad (85b)$$

This wave function and the corresponding wave intensity are really well localised around the object coordinate eigenvalue $x = X_i$, the fact that is *deduced* from the general expressions of our formalism without any supplementary assumptions. Moreover, for this example of coordinate measurement we can trace also partial localisation of $\Psi_i(x,X)$ and $\rho_i(x,X)$ in the instrument configurational space, around $X = X_i$. It follows directly from the expression for the effective interaction operator, eqs. (76), the properties of matrix elements $V_{0i}(X)$, eq. (69), and the expression for $\eta_{1,2n}^i(X,X')$, eq. (84) (see also Fig. 4): all those functions of $X$ have much larger magnitude at $X \approx X_i$, for the $i$-th realisation (it is verified by substitution of the values of $\chi_n^i$, Fig. 4, into eqs. (76), (84)). This causal wave reduction is quite consistent with, and naturally linked to, the above indeterminacy: as the projectile $i$-th realisation is due mainly to the interaction with the $i$-th detector, and the realisations cannot coexist, then each of them is naturally 'concentrated' around 'its' detector.

One should take into account that quantitatively the reduction and localisation effects described are much more pronounced in the real situation, where one has a great number of very closely spaced eigenvalues $x'$, $X_g$, and $\chi_n$. As it is evident from the above analysis (and also from that for the general case, section 9.1), the characteristic size of the dynamically reduced wave, $\Delta\chi_n$, is determined by the separation of the measured quantity eigenvalues, $\Delta x = X_2 - X_1$ in the present model. The weak, but still well-defined, actual localisation is due to the extremely large eigenvalue separation, the unavoidable price for the dynamical simplicity of the model. In real coordinate measurement process, $\Delta x$ cannot be larger than the detector interatomic spacing, and eventually this minimum coordinate separation should probably be still much smaller, of the order of the characteristic extension, now unknown, of the de Broglie quasi-soliton.

The presented detailed description of the configurational wave-function reduction gives rise to another, though related and equivalent, interpretation of the occurring process in simple physical terms. Namely one may speak of the intrinsic *instability* of the system {projectile + detectors} with respect to the existing (three) possibilities of projectile interaction with each of the detectors (and the obstacle). For the really linear wave this instability could not appear and the wave would coherently and symmetrically interact with all parts of the instrument. The key point of our analysis is that it shows why and how precisely the *apparent* Schrödinger wave linearity turns into its *intrinsic effective nonlinearity*, for any system with many interacting degrees of freedom.

To express it in qualitative physical terms, note, for example, that each of two arbitrary neighbouring points (small volumes) of the measured wave interacts with each of the degrees of freedom of the instrument (two detectors and the obstacle in our current model). But this means that those two points are no more independent and should show certain effective interaction between them. Such self-interacting wave evidently gains nonlinearity, just because of its interaction with the instrument. This effective nonlinearity has been simply hidden in the ordinary, non-modified description, eq. (65), and it explicitly appears in the modified form of the same equations, eq. (75) (cf. the same transformation in the general theory of measurement, section 9.1, and also in the Hamiltonian quantum chaos description, parts I-III).[*] The effectively nonlinear wave shows typical instability with respect to the localised structure formation which can equally well happen around each of the interaction centers. These localised nonlinear forms, exemplified by eqs. (84), (85), can no more be coherent, neither can they coexist: a more successive participant, randomly

---

[*] By the way, it provides also a general definition of such *intrinsic* nonlinearity which is inseparably related to the fundamental dynamic multivaluedness and complexity (section 6.III) and differs substantially from the usually employed *formal* nonlinearity, much more ambiguous and compatible, for example, with some *exact* solutions (zero complexity), e. g. solitons.

'selected' by local internal excitations (i. e. eventually by small fluctuations), will always 'eat up' the others. In our formal description this behaviour corresponds to the existence of *many complete* realisations.

We shall finally use matching conditions at the boundary surface ($z = 0$) to determine the coefficients $c_n^i$ in eq. (85a). Using eq. (64) we obtain the boundary condition for the $i$-th realisation in the form

$$\Phi_0(x) \sum_{n,n'} c^0_{nn'} \delta(X - X^0_{nn'}) = \sum_{n=1}^{2} c_n^i \delta(x - X_i) \hat{\eta}_{in}^i(X) \psi_{0n}^i(X) \ . \tag{86}$$

Multiplying it by $\delta(X_i - x)$ and integrating over $X$ and $x$, we obtain

$$\Phi_0(X_i) = \sum_{n=1}^{2} c_n^i \equiv C_i \ . \tag{87}$$

This result has the fundamental physical meaning: it can be interpreted as the conventional quantum-mechanical rule of probabilities, but now *deduced* as a consequence of the fundamental dynamic uncertainty which is also naturally obtained from the effective measurement equation. Indeed, averaging eq. (86) over the instrument coordinates, $X$, corresponds to smearing of the reduced-wave intensity peaks of the individual registration events. What remains then is the distribution of probability (or density, for many events) of occurrence of these peaks-realisations, eq. (87). Namely, the probability of the $i$-th realisation, $\alpha_i \equiv |C_i|^2$, appearing in the close vicinity of $x = X = X_i$, is equal to $|\Phi_0(X_i)|^2 \equiv |a_{x=X_i}|^2$ (see eq. (62c)). For a realistic situation with many close realisations, this means that $|\Phi_0(x)|^2$ determines the *probabilistic density of realisations* around $x$ (cf. eqs. (16c.II)). We thus confirm and specify the fundamental *dynamic* origin of the irreducible probabilistic element in quantum-mechanical postulates: it is the fundamental uncertainty of the indefinite choice within the redundant set of realisations (cf. the general analysis of section 9.1).

We have just specified the realisation probabilities, $\{\alpha_i\}$. As those probabilities, according to the dynamic uncertainty postulate (see e. g. eq. (16.II)), should be independent from the internal structure of realisations, it is natural to assume, based on the definition, eq. (87), that

$$c_n^i = C_i c_n \ , \quad \sum_{n=1}^{2} c_n = 1 \ . \tag{88}$$

The coefficients $c_n$ characterise the details of the *deterministic* density distribution within each event-realisation and can be determined from the unreduced boundary condition, eq. (86). However, these details are practically of much less interest (in reality the individual events can be considered most often as point-like ones) and we shall not specify them (see eq. (58b) in the general description and the accompanying discussion). For the same reason we do not analyse possible weak dependence of these reduced coefficients $c_n$ on the realisation number $i$.

The performed analysis can be directly extended to another formulation of the same problem. For the situation of Fig. 3, one may be interested in the second measurement which happens when a projectile, after passage through the holes, creates a registration event in the sensitive screen. For the particular case when a wave attaining the screen is close to the plane wave, we obtain the classic formulation of the problem of wave-particle duality which was actively employed, among others, by Einstein already in the early years of quantum mechanics for a transparent 'physical' demonstration of its incompleteness.

The screen is reasonably approximated by the crystal lattice of atoms which can be individually excited by the projectile, this leading eventually to the individual registration events. Each atom can be considered then as an elementary detector in our previous scheme. It is then clear that the problem is an extension of the preceding one with two detectors to the large number, $N_a \gg 1$, of detector-atoms arranged in a flat lattice (the case of three-dimensional lattice of atoms can also be included without any major changes). The generalisation of the results obtained is thus quite evident; in the expressions above $X_n$ will designate the screen atom positions, while the subscript $g$, enumerating the measured eigenvalues, takes $N_a$ values. It remains then simply to reproduce the above analysis, with only minor technical changes, to obtain practically the same causal description of quantum-mechanical indeterminacy and wave reduction involved in this long-standing problem of projectile impinging on a screen. This description provides the clear solution to quantum paradoxes formulated in the historical critics of the standard interpretation and should eventually be completed by the *almost independent* nonlinear local picture of wave mechanics (see section 10.V and e-prints quant-ph/9902015,16).

While there is no sense to repeat in detail the same results that have just been specified for the case of two detectors, we can very briefly summarise the fundamental physical essence of our solution to the quantum mysteries. Namely, we have shown that this 'magic' source of the peculiar wave-particle transformations that was tentatively attributed either to some hidden variables and even reality levels, or to the unusual logic of quantum theory itself, can be consistently and universally presented as the true dynamical randomness in the non-Hamiltonian quantum systems with (basically) few degrees of freedom. We have seen that it is this general mechanism of the fundamental dynamic multivaluedness, involving both the deterministic origin and the *deduced* probabilistic character of the dynamic uncertainty, that can successfully replace the inconsistent reference to particular "classical", or "macroscopic", nature of the instrument playing the crucial role in the conventional measurement descriptions. Full quantum indeterminacy and reduction are now permitted within quite microscopic, and essentially quantum, systems which should only possess more than one degrees of freedom and among them at least one

'excitation' degree creating an open system.*⁾ It is important that those additional 'external' degrees of freedom remaining beyond the main description are not involved with the fundamental irreversibility of the measurement process; the latter is a natural manifestation of no less fundamental dynamical indeterminacy.

The results of the performed analysis, in section 9, of the fundamental dynamic uncertainty involvement in the basic problems of quantum mechanics can now be summarised within a further modification of our tentative answers (36) to the basic questions (35):

$$\begin{array}{r}\text{Quantum mechanics (in the modified form)}\\ \text{obeys the correspondence principle.}\\ \text{It is formally complete.}\end{array} \qquad (89)$$

Now we are going to show, in the next section, why and how these answers can be yet more extended to allow for the *physically* complete scheme of quantum mechanics.

---

*⁾ The possibility of practical realisation of the exotic situation of excitable measurement system with only one effective degree of freedom, providing the unusual completely deterministic behaviour, is a questionable issue and could be tempted as an interesting particular case that may somehow complete the general results.